\documentclass[a4paper,twoside]{article}

\usepackage{epsfig}
\usepackage{subcaption}
\usepackage{calc}
\usepackage{amssymb}
\usepackage{amstext}
\usepackage{amsmath}
\usepackage{amsthm}
\usepackage{multicol}
\usepackage{pslatex}
\usepackage{apalike}
\usepackage{algorithm2e}
\usepackage[bottom]{footmisc}
\usepackage{caption}
\usepackage{csvsimple}
\usepackage{threeparttable}
\usepackage{makecell}
\usepackage{diagbox}
\usepackage{booktabs, makecell}
\usepackage{listings}
\usepackage[table]{xcolor}
\usepackage{textcomp}
\usepackage{array}
\usepackage{pgfplots}
\usepackage{enumitem}
\usepackage[hidelinks]{hyperref}
\usepackage{cleveref}
\usepackage{multirow} 
\usepackage{comment}
\usepackage{titlesec}

\usepackage{SCITEPRESS}     

\newcommand{\PreserveBackslash}[1]{\let\temp=\\#1\let\\=\temp}
\newcolumntype{C}[1]{>{\PreserveBackslash\centering}p{#1}}
\newcolumntype{R}[1]{>{\PreserveBackslash\raggedleft}p{#1}}
\newcolumntype{L}[1]{>{\PreserveBackslash\raggedright}p{#1}}
\newcolumntype{x}[1]{>{\centering\arraybackslash\hspace{0pt}}p{#1}}

\definecolor{codegreen}{rgb}{0,0.6,0}
\definecolor{codegray}{rgb}{0.5,0.5,0.5}
\definecolor{codepurple}{rgb}{0.58,0,0.82}
\definecolor{backcolour}{rgb}{0.98,0.98,0.96}

\lstdefinelanguage{BPPy}
{morekeywords={interrupt, block, request, waitFor, Block, Request, mustFinish, localReward, yield, while, def, for, in, range, from, import, return, if, else, lambda, sum, filter, class, super},
sensitive=false,
morecomment=[l]{\#},
morecomment=[s]{/*}{*/},
morestring=[b]",
morestring=[b]',
}

\lstdefinestyle{mystyle}{
    language=BPPy,
    backgroundcolor=\color{backcolour},   
    commentstyle=\color{codegreen},
    keywordstyle=\color{blue},
    keywordstyle=[5]\ttfamily\scriptsize
    numberstyle=\tiny\color{codegray},
    stringstyle=\color{codepurple},
    basicstyle=\ttfamily\tiny,
    breakatwhitespace=false,         
    breaklines=true,                 
    captionpos=b,                    
    keepspaces=true,                 
    numbers=none,                    
    numbersep=5pt,                  
    showspaces=false,                
    showstringspaces=false,
    showtabs=false,                  
    tabsize=2
}

\lstdefinestyle{mystyle2}{
    language=BPPy,
    backgroundcolor=\color{backcolour},   
    commentstyle=\color{codegreen},
    keywordstyle=\color{magenta},
    keywordstyle=[5]\ttfamily\scriptsize
    numberstyle=\tiny\color{codegray},
    basicstyle=\ttfamily\scriptsize,
    breakatwhitespace=false,         
    breaklines=true,                 
    captionpos=b,                    
    keepspaces=true,                 
    numbers=none,                    
    numbersep=5pt,                  
    showspaces=false,                
    showstringspaces=false,
    showtabs=false,                  
    tabsize=2
}

\newcommand{\li}[2][]{\lstinline[basicstyle=\ttfamily\normalsize,breaklines=true,prebreak=,#1]{#2}}

\newtheorem{definition}{Definition}
\newtheorem{example}{Example}

\pgfplotsset{compat=1.7}

\begin{document}

\title{Exploring and Evaluating Interplays of BPpy with \\ Deep Reinforcement Learning and Formal Methods\vspace{-0.1cm}}

\author{\authorname{Tom Yaacov\sup{1}, Gera Weiss\sup{1}, Adiel Ashrov\sup{2}, Guy Katz\sup{2}, and Jules Zisser\sup{1}}
\affiliation{\sup{1}Ben-Gurion University of the Negev}
\affiliation{\sup{2}The Hebrew University of Jerusalem}
\email{tomya@post.bgu.ac.il, geraw@bgu.ac.il, adiel.ashrov@mail.huji.ac.il, guykatz@cs.huji.ac.il, zisserh@post.bgu.ac.il}
\vspace{-1.1cm}
}

\keywords{Behavioral Programming, Deep Reinforcement Learning, Formal Methods\vspace{-1cm}}

\abstract{We explore and evaluate the interactions between Behavioral Programming (BP) and a range of Artificial Intelligence (AI) and Formal Methods (FM) techniques. Our goal is to demonstrate that BP can serve as an abstraction that integrates various techniques, enabling a multifaceted analysis and a rich development process. Specifically, the paper examines how the BPpy framework, a Python-based implementation of BP, is enhanced by and enhances various FM and AI tools. We assess how integrating BP with tools such as Satisfiability Modulo Theory (SMT) solvers, symbolic and probabilistic model checking, and Deep Reinforcement Learning (DRL) allow us to scale the abilities of BP to model complex systems. Additionally, we illustrate how developers can leverage multiple tools within a single modeling and development task. The paper provides quantitative and qualitative evidence supporting the feasibility of our vision to create a comprehensive toolbox for harnessing AI and FM methods in a unified development framework.
\vspace{-0.3cm}}

\onecolumn \maketitle \normalsize \setcounter{footnote}{0} \vfill

\section{Introduction}
\label{sec:introduction}
\vspace{-3pt}

It is commonly agreed that the future of software development, especially in reactive systems, relies on the use of models, advanced analysis techniques, and artificial intelligence (AI)~\cite{MDE24}. However, while the current state-of-the-art involves various analysis tools and techniques, each still uses its own input language and modeling approach, requiring manual integration to combine results with system code. Finding a modeling framework that can safely interweave user code and machine-generated artifacts while supporting formal analysis is a challenging research problem. This paper examines the Behavioral Programming (BP) framework \cite{harel_behavioral_2012} and its potential as a unified modeling abstraction that glues tools without manual translation, showing its effectiveness in combining multiple tools throughout the software development process.

BP is a software engineering paradigm designed to allow developers to specify the behavior of reactive systems incrementally and intuitively, aligning with how they perceive the system's requirements~\cite{harel_behavioral_2012}. Its primary strength lies in its ability to break down intricate specifications into manageable components that interact through a unified protocol, enabling the creation of desired behavior. This compositional modeling capability can be applied across various domains~\cite{bar-sinai_bpjs_2018} and integrated with techniques from different disciplines~\cite{bar-sinai_verification_2021}.

The benefits afforded by BP to traditional software engineering have been studied extensively~\cite{elyasaf_using_2019,elyasaf_context-oriented_2021}. In this work, we aim to complement existing research by focusing on the problem at hand --- namely, the integration of BP with AI and formal analysis (FM) techniques. Specifically, we utilize the BPpy library~\cite{yaacov_bppy_2023}, a framework for BP in Python, along with several Python-based libraries, to explore the interplay and benefits that arise from these combinations. Ultimately, we aim to lay the groundwork for constructing a comprehensive software engineering toolbox to harness modern AI and FM techniques.

Our contributions include:

\begin{description}[leftmargin=0cm]
\item[\emph{Satisfiability-Modulo-Theory (SMT):}] We present an enhanced communication protocol among modules for solving complex constraint systems and show it promotes the efficiency of the execution mechanism. 
\item[\emph{Symbolic model checking:}] We developed a method that allows the verification of large BP models by analyzing each module separately and combining them symbolically. 
\item[\emph{Probabilistic model checking:}] We designed an approach that facilitates the analysis of probabilistic aspects of a behavioral system.
\item[\emph{Deep Reinforcement Learning (DRL):}] We present a framework that improves the alignment of requirements with BP modules, delivering a more efficient execution mechanism.
\end{description}
 
While this is not the main focus of the paper, it is worth noting that benefits go both ways: BP's effective system components management can boost these tools. 
For instance, BP can serve as a theory for SMT solvers by dynamically generating candidate runs and incrementally revealing constraints, like other theories guide the solver's reasoning process.
In the context of \emph{DRL}, we leverage the flexibility offered by BP to specify a system's behavior only partially, leaving gaps for the DRL engine to resolve. This approach can help guide the learning process and safeguard the resulting execution mechanism from leading to undesirable outcomes. BP's flexible division of the model into smaller modules can also aid in \emph{symbolic or probabilistic verification} process and facilitate more natural and direct modeling. This differs from current alternatives, focusing on a monolithic state-based system description.

This study not only focuses on new integrations. While the symbolic and probabilistic verification integrations presented here are novel, the combination of SMT and DRL with BP was discussed in prior research \cite{katz_--fly_2019,eitan_adaptive_2011,elyasaf_using_2019,yaacov_bppy_2023}. To provide a comprehensive understanding of the benefits of these integrations, we incorporated ideas from previous work, introduced new tools, and conducted fresh evaluations comparing them against current alternatives in BP to assess the effectiveness of all tools in new dimensions.

%


%

As a glimpse into a future research direction, we also present a small experiment showing the applicability of combining SMT solvers, probabilistic modeling, and DRL in conjunction with BP. The result of this study provides compelling evidence, in our opinion, of the viability of using BP as a modeling approach that supports all the aforementioned methods.

The structure of the paper is as follows: We start with a short introduction to BP and BPpy in \autoref{sec:bppy}. Subsequently, each integration is discussed individually in \cref{sec:smt,sec:model-checking,sec:probabilistic-modeling,sec:drl}. These sections include evaluations and analyses. We do not devote separate sections to related work and comprehensive conclusions, as relevant references and discussion are provided within each section. In \autoref{sec:four-way-integration}, we showcase the combined use of these integrations. All supplementary material is available at { \url{https://github.com/bThink-BGU/Papers-2025-ENASE-BPpyEvaluation}}.

\vspace{-10pt}
\section{Behavioral Programming \& BPpy}
\label{sec:bppy}
\vspace{-3pt}

To introduce the reader to the language of BP, we begin by describing the BPpy package and the general BP principles it implements. In BP, developers specify scenarios, named \emph{b-threads}, which are simple sequential threads of execution that represent behaviors the system should include or avoid. Each scenario is standalone and is usually self-contained, concerning itself with a specific aspect of the system---typically, a single requirement. During runtime, an application-agnostic execution mechanism interprets and interweaves these b-threads to generate a cohesive system behavior. Specifically, the mechanism is based on a synchronization protocol presented by~\cite{harel_programming_2010}. It consists of each b-thread submitting a statement before selecting each event that the b-program produces. When a b-thread reaches a point where it is ready to submit a statement, it synchronizes a statement with its peers, declaring which events it \emph{requests}, which events it \emph{waits for} (but does not request), and which events it \emph{blocks} (prevents from occurring). After submitting the statement, the b-thread is paused. Once all b-threads have submitted their statements, we say the b-program has reached a \emph{synchronization point}. At this point, an event arbiter selects a single event that has been requested and is not blocked. It then resumes all b-threads that either requested or waited for the selected event. The other b-threads remain paused, and their statements are used in the next synchronization point. This process is repeated throughout the execution of the program. A formal definition of BP semantics is available as an appendix in the supplementary material.

To make BP's core concepts more concrete, we begin with an illustrative example of a \emph{b-program} (a set of b-threads) implemented in BPpy. The example is an adaptation of one of the sample b-programs presented in the work of~\cite{harel_behavioral_2012} describing a system responsible for controlling the mixing of hot and cold water from two separate taps. \autoref{lst:hot-cold-intro} depicts three b-threads. Each b-thread is implemented as a Python generator---a function that can pause itself and pass data back to its caller at any point, using the \li{yield} idiom. It can then be resumed when re-invoked with the \li{send} method. The statements submitted by each b-thread are structured as \li{sync} class instances containing events or event sets labeled by the arguments \li{request}, \li{block}, \li{waitFor}. The execution mechanism in BPpy starts by independently invoking each b-thread generator and awaiting its statement yield. Once all the statements have been collected, an event is selected, and the program resumes its execution based on the aforementioned synchronization protocol.

\begin{lstlisting}[
label={lst:hot-cold-intro},
caption={The HOT/COLD b-program~\cite{harel_behavioral_2012}.},
float=htb
]
@thread
def add_hot():
  for i in range(3):
    yield sync(request=BEvent("HOT"))

@thread
def add_cold():
  for i in range(3):
    yield sync(request=BEvent("COLD"))

@thread
def control():
  while True:
    yield sync(waitFor=BEvent("HOT"))
    yield sync(waitFor=All(), block=BEvent("HOT"))
\end{lstlisting}

The first two b-threads, \li{add_hot} and \li{add_cold}, request the event of pouring a small amount of hot and cold water, respectively, three times. Unlike many other programming paradigms, BP offers developers the flexibility not to be bound by a single predefined behavior for the system. Instead, the system has the freedom to select any behavior that aligns with all the defined b-threads. For instance, a b-program consisting of the two b-threads shown in \autoref{lst:hot-cold-intro} does not impose a specific order for pouring hot and cold water. Consequently, its execution can generate all sequences that include exactly three occurrences of the \li{HOT} event and three occurrences of the \li{COLD} event. 

To illustrate further, consider that after running the initial version of the system for some time, a safety concern arises, and a new requirement is introduced, stating that it is undesirable to have two consecutive \li{HOT} events. While it is possible to modify the \li{add_hot} b-thread by adding new conditions and statements, the BP paradigm encourages us to maintain the alignment between existing b-threads and their respective requirements and add a new b-thread. This approach promotes an incremental and modular development style, where developers can add or remove behaviors independently without affecting other b-threads. Thus, we introduce the \li{control} b-thread in \autoref{lst:hot-cold-intro}, which repeatedly waits for the occurrence of \li{HOT} and then blocks \li{HOT}  while waiting for any following event using the \li{All} event set.

\vspace{-10pt}
\section{BP \texorpdfstring{$\Leftrightarrow$}{<->} SMT Solvers}
\label{sec:smt}
\vspace{-3pt}
BPpy implements SMT solver integration following the concepts outlined in~\cite{katz_--fly_2019}. The implementation is based on the Z3-solver~\cite{de_moura_z3_2008} package, although other solvers may also be used. In this integration, events are represented as an assignment over a set of SMT variables. At each synchronization point, b-threads specify \li{request}/\li{block}/\li{waitFor} \emph{constraints} over the variables in the form of logical statements to be satisfied. Once all constraints are collected, the execution mechanism invokes the solver to find a satisfying assignment to the variables embedded in these constraints. Specifically, the solver finds an assignment requested by at least one b-thread that is not blocked. 

To introduce this integration, we begin with an illustration of the \emph{Cinderella-Stepmother} problem~\cite{bodlaender2012cinderella}. This problem involves a two-player game with a system of water buckets. Initially, there are \li{N} empty buckets arranged in a circle, each with a capacity of \li{B} water units. In each turn, Cinderella's stepmother distributes \li{A} water units across the buckets as she chooses. Subsequently, Cinderella empties \li{C} adjacent buckets. This cycle of the stepmother pouring and Cinderella emptying repeats. The stepmother's objective is to fill one bucket with \li{B} units, while Cinderella aims to prevent any bucket from becoming full.

The code in \autoref{lst:smt1} exemplifies a solver-based version of the Cinderella-Stepmother example. It consists of the \li{main} b-thread, which is responsible for changing the assignments to the \li{buckets} list of integers based on the constraints generated by the \li{stepmother} and \li{cinderella} functions. These functions use the last selected assignment, which is returned through the b-thread's \li{yield} command. The \li{bucket\_limit} b-thread ensures that the variables will not exceed the buckets' capacity \li{B}. We note that  \li{bucket\_limit} does not specify request or wait-for constraints. Thus, its blocking constraint remains invariant throughout the execution of the b-program.

\begin{lstlisting}[
label={lst:smt1},
caption={A BP solver-based implementation to the Cinderella-Stepmother program.},
float=htb
]
buckets = [Int(f"b{i}") for i in range(N)]

def stepmother(prev):
 added=Sum([b-prev.eval(b) for b in buckets])
 non_neg=And([b-prev.eval(b)>=0 for b in buckets])
 return And(added == A,non_neg)

def cinderella(prev):
 r = list(range(N)) + list(range(N))
 def empty(rng):
  cs = [] # constraints list
  for j in range(N):
   if j in rng:
    cs.append(buckets[j] == 0)
   else:
    cs.append(buckets[j] == prev.eval(buckets[j])) 
  return And(cs)
 return Or([empty(r[i:i+C]) for i in range(N)])

@thread
def bucket_limit():
 while True:
  yield sync(block=Or([b > B for b in buckets]))

@thread
def main():
 e=yield sync(request=And([b==0 for b in buckets])
 for i in range(STEPS):
  e = yield sync(request=stepmother(e))
  e = yield sync(request=cinderella(e))
\end{lstlisting}

The SMT solver integration allows b-threads to communicate versatile variable assignments, going beyond discrete events. Along with its advantages in enhancing expressiveness, extensively discussed in~\cite{katz_--fly_2019}, it can potentially improve BP's computational capabilities during execution. This section aims to validate this assumption by comparing the conventional discrete event mechanism and the presented SMT-based event mechanism. To assess this, a set of empirical experiments was carried out to measure the runtime execution and memory efficiency of a b-program tasked with solving a multiple-constraint problem using both approaches. The complete code developed for this evaluation is available in the supplementary material.

The evaluation included three use cases. The first is the Cinderella-Stepmother problem, for which we developed an equivalent discrete implementation and compared its performance with the solver-based implementation presented in \autoref{lst:smt1}. This implementation generates all the possible states of the buckets as individual events. We tested the programs with increasing parameter values of \li{B} and \li{N}. Since the complexity increases very rapidly in the discrete case with \li{N}, we only show here the growth with \li{B}. 

The second use case is an adaptation of the \emph{Lights Out} puzzle game~\cite{anderson1998turning}, which we refer to as the \emph{bit-flip} problem. The problem domain is a Boolean matrix of dimensions $N\times M$. At first, the matrix is randomly initialized, and in each move, a single row or column values are flipped. Also, in each move, one row or column cannot be flipped, and it is therefore blocked. A potential objective is to find a sequence of bit-flips to transition the matrix from one configuration to another. This problem is relevant as Boolean matrices can effectively represent reactive systems such as communication networks.

\begin{figure}[h]
  \begin{minipage}[c]{0.27\textwidth}
    \resizebox{0.90\textwidth}{!}{%
        \begin{tikzpicture}
            \draw (-3,0) -- (0,3) -- (3,0) -- (0,-3) -- cycle;

            \draw[black, line width=2pt] (0,0) circle [radius=3];

            \draw[->] (-3.5,0) -- (3.5,0) node[below] {$x$};
            \draw[->] (0,-3.5) -- (0,3.5) node[left] {$y$};

            \node[below, rotate=45] at (-1.5,1.5) {$y = x + 1$};
            \node[below, rotate=315] at (1.5,1.5) {$y = -x + 1$};
            \node[above, rotate=45] at (1.5,-1.5) {$y = x - 1$};
            \node[above, rotate=315] at (-1.5,-1.5) {$y = -x - 1$};

            \node [below, rotate=45, text=black] at (-2.6,2.6) {$x^2+y^2=1$};

            \fill[blue!50, even odd rule] (-3,0) -- (0,3) -- (3,0) -- (0,-3) -- cycle (0,0) circle [radius=3];

            \fill (3,0) circle [radius=3pt] node[above right] {$(1,0)$};
            \fill (0,3) circle [radius=3pt] node[above right] {$(0,1)$};
            \fill (0,0) circle [radius=3pt] node[above right] {$(0,0)$};
            \fill (-3,0) circle [radius=3pt] node[above left] {$(-1,0)$};
            \fill (0,-3) circle [radius=3pt] node[below right] {$(0,-1)$};
        \end{tikzpicture}%
    }
  \end{minipage}\hfill
  \begin{minipage}[c]{0.18\textwidth}
    \caption{
       A circled polygon problem with 4 edges. The colored area between the polygon and the circle represents the area of potential solutions to the problem.
    } \label{fig:regular_polygon_multi_edge}
  \end{minipage}
\end{figure}
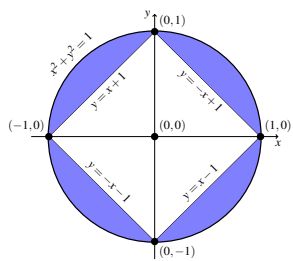

For the third example, we implemented the \emph{circled polygon} example (see \autoref{fig:regular_polygon_multi_edge}). The program's objective is to find a coordinate outside a regular polygon's area but inside the circumscribing unit circle. We tested this problem with an increasing number of polygon edges. The problem relevance stems from its multifaceted nature, common in reactive systems with geometrical constraints involving real-valued parameters that BP is well-suited to solve~\cite{katz_--fly_2019,elyasaf_context-oriented_2021}. The discrete implementation solves this multi-constraint problem by discretizing the product of the two-dimensional continuous intervals $[-1,1]^2$ (the area enclosing the circle centered at (0, 0)) incrementally. As the number of edges in the polygon grows, the area between the polygon and the circle diminishes, necessitating finer discretization and an increase in the number of events to find a solution.

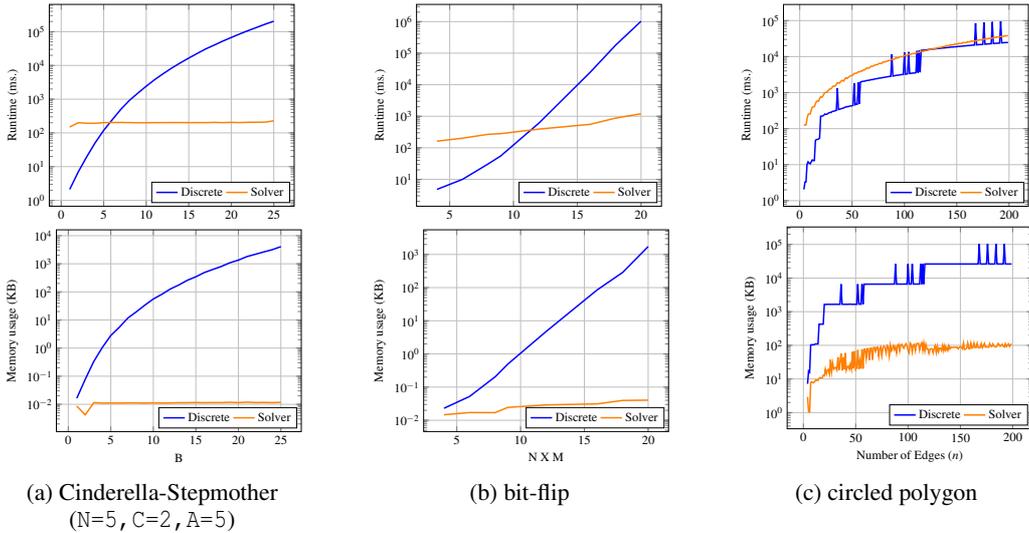
\begin{figure*}[hbt]
    \centering
    \captionsetup[subfigure]{justification=centering}
        \begin{subfigure}[t]{0.3\textwidth}
               \centering
        \begin{tikzpicture}[scale=0.47]
        \begin{axis}[
            ylabel={Runtime (ms.)},
            grid=major,
            legend style={at={(0.7,0.12)}, anchor=north, legend columns=-1},
            ymode=log,
            log basis y=10,
        ]
        \addplot [color=blue, very thick] table [x=B, y=execution_time_discrete, col sep=comma] {cinderella_time.csv};
        \addlegendentry{Discrete}
        
        \addplot [color=orange, very thick] table [x=B, y=execution_time_smt, col sep=comma] {cinderella_time.csv};
        \addlegendentry{Solver}
        
        \end{axis}
        \end{tikzpicture}
        \label{fig:cinderella_runtime_comparison}
        \end{subfigure}
        \begin{subfigure}[t]{0.3\textwidth}
               \centering
        \begin{tikzpicture}[scale=0.47]
        \begin{axis}[
            ylabel={Runtime (ms.)},
            grid=major,
            legend style={at={(0.7,0.12)}, anchor=north, legend columns=-1},
            ymode=log,
            log basis y=10,
        ]
        \addplot [color=blue, very thick] table [x=n_times_m, y=execution_time_discrete, col sep=comma] {bit_flip_time.csv};
        \addlegendentry{Discrete}
        
        \addplot [color=orange, very thick] table [x=n_times_m, y=execution_time_smt, col sep=comma] {bit_flip_time.csv};
        \addlegendentry{Solver}
        
        \end{axis}
        \end{tikzpicture}
        \label{fig:bit_flip_runtime_comparison}
        \end{subfigure}
        \begin{subfigure}[t]{0.3\textwidth}
                \centering
        \begin{tikzpicture}[scale=0.47]
        \begin{axis}[
            ylabel={Runtime (ms.)},
            grid=major,
            legend style={at={(0.7,0.12)}, anchor=north, legend columns=-1},
            ymode=log,
            log basis y=10
        ]
        \addplot [color=blue, very thick] table [x=num_of_edges, y=execution_time_discrete, col sep=comma] {multi_edge_time.csv};
        \addlegendentry{Discrete}

        \addplot [color=orange, very thick] table [x=num_of_edges, y=execution_time_solver, col sep=comma] {multi_edge_time.csv};
        \addlegendentry{Solver}
        
        \end{axis}
        \end{tikzpicture}
        \label{fig:multi_edge_runtime_comparison}
        \end{subfigure}
        \hfill
        \begin{subfigure}[t]{0.3\textwidth}
               \centering

        \begin{tikzpicture}[scale=0.47]
        \begin{axis}[
            xlabel={B},
            ylabel={Memory usage (KB)},
            grid=major,
            legend style={at={(0.7,0.12)}, anchor=north, legend columns=-1},
            ymode=log,
            log basis y=10
        ]
        \addplot [color=blue, very thick] table [x=B, y=memory_usage_discrete, col sep=comma] {cinderella_memory.csv};
        \addlegendentry{Discrete}
        
        \addplot [color=orange, very thick] table [x=B, y=memory_usage_smt, col sep=comma] {cinderella_memory.csv};
        \addlegendentry{Solver}
        
        \end{axis}
        \end{tikzpicture}
        \caption{Cinderella-Stepmother (\li{N=5,C=2,A=5})}
        \label{fig:cinderella_memory_comparison}
        \end{subfigure}
        \begin{subfigure}[t]{0.3\textwidth}
               \centering

        \begin{tikzpicture}[scale=0.47]
        \begin{axis}[
            xlabel={N X M},
            ylabel={Memory usage (KB)},
            grid=major,
            legend style={at={(0.7,0.12)}, anchor=north, legend columns=-1},
            ymode=log,
            log basis y=10
        ]
        \addplot [color=blue, very thick] table [x=n_times_m, y=memory_usage_discrete, col sep=comma] {bit_flip_memory.csv};
        \addlegendentry{Discrete}
        
        \addplot [color=orange, very thick] table [x=n_times_m, y=memory_usage_smt, col sep=comma] {bit_flip_memory.csv};
        \addlegendentry{Solver}
        
        \end{axis}
        \end{tikzpicture}
        \caption{bit-flip}
        \label{fig:bit_flip_memory_comparison}
        \end{subfigure}
        \begin{subfigure}[t]{0.3\textwidth}
                \centering
        \begin{tikzpicture}[scale=0.47]
        \begin{axis}[
            xlabel={Number of Edges ($n$)},
            ylabel={Memory usage (KB)},
            grid=major,
            legend style={at={(0.7,0.12)}, anchor=north, legend columns=-1},
            ymode=log,
            log basis y=10
        ]
        \addplot [color=blue, very thick] table [x=num_of_edges, y=memory_usage_discrete, col sep=comma] {multi_edge_memory.csv};
        \addlegendentry{Discrete}

        \addplot [color=orange, very thick] table [x=num_of_edges, y=memory_usage_solver, col sep=comma] {multi_edge_memory.csv};
        \addlegendentry{Solver}
        
        \end{axis}
        \end{tikzpicture}
        \caption{circled polygon}
        \label{fig:multi_edge_memory_comparison}
        \end{subfigure}
        \caption{
    The runtime and memory of the discrete and solver-based implementations for the three problems. The values on the y-axis were converted to $log_{10}$ scale due to the variance in the original scales.}
    \label{fig:smt-results}
\end{figure*}

The memory and runtime results for the three use cases are shown in \autoref{fig:smt-results}. For the Cinderella-Stepmother and bit-flip problem, the runtime and memory performance of the discrete implementation exponentially increases with \li{B} and $N \times M$, respectively. This behavior aligns with expectations, as increasing parameter values increase the number of events, which means that more events are being examined at the synchronization point, and hence, additional runtime and memory are needed. The runtime and memory usage of the solver-based Cinderella-Stepmother implementation remained constant as the problem size increased, while the solver-based bit-flip implementation increased exponentially but with a significantly smaller slope.

For the circled polygon problem, both implementations exhibit a significant increase in runtime as the complexity rises, contrasting with previous results. We attribute this latter increase to the fact that we now have many equations over real-valued variables; this may be because, in this model, the SMT engine runs the theory module many times. Interestingly, the runtime of the discrete implementation remains similar or lower until $n=164$, after which it spikes significantly. This behavior stems from varying complexities, where some instances are straightforward for the discrete implementation, while others require finer discretization. Regarding memory usage, the solver-based implementation maintains a distinct advantage, stabilizing around $n=100$. Nevertheless, the memory values surpass those of previous examples, underscoring the problem's complexity for the solver. Conversely, the discrete version memory usage increases with finer discretization as complexity rises, with spikes similar to the runtime analysis, indicating the discussed unexpected behavior.

In summary, our study encompassed a performance evaluation contrasting discrete and solver-based implementations across the domains of Boolean, Integer, and Real variables, with various types of constraints and different levels of complexities. The experiments confirmed the anticipated superiority of the solver-based approach. Additionally, we showcased the adaptability of BP's SMT solvers integration in tackling various problem scenarios.

The potential use of BP to enhance SMT capabilities has been explored in prior studies~\cite{harel_composing_2013}. While SMT solvers are crucial for various applications like software verification and program synthesis, their incompleteness can hinder their effectiveness. Integrating BP with SMT can enhance system robustness and reliability, potentially leading to more advanced automated theorem-proving tools for specific theories like arithmetic and arrays. As mentioned in the introduction, BP can serve as a theory for deciding behavioral constraints in a setting where we try to find runs that satisfy a set of constraints modeled as a composition of b-threads. 

Qin et al. utilized SMT in conjunction with BIP models, employing it to prune unsatisfiable transitions in Open Automata derived from Open pNets~\cite{qin2020smt}. Their work emphasizes formal analysis rather than enhancing the underlying mechanism. In contrast, our approach integrates SMT into BP to introduce a decision mechanism for dynamically selecting events, thereby increasing its expressiveness and adaptability.

\vspace{-10pt}
\section{BP \texorpdfstring{$\Leftrightarrow$}{<->} Symbolic Model Checking}
\label{sec:model-checking}
\vspace{-3pt}
This section introduces a new method for verifying b-programs by translating them into the SMV symbolic specification language~\cite{mcmillan_smv_1993}. A significant advantage of this method is its ability to verify software written in languages like Python, where the states of the b-threads, modeled as generators, are not readily clonable~\cite{yaacov_bppy_2023}. We evaluate the efficiency of symbolic model checking in BP programs and compare it with the current best practices in BP verification, which rely on explicit model checking, involving an exhaustive enumeration of the state space. 

As in other modeling paradigms, verification is a subject of considerable work. 
Some introduced model checking methodologies and accompanying tools for verifying safety and liveness properties in behavioral programs~\cite{harel_model-checking_2011,bar-sinai_verification_2021}.
Other studies have explored the potential benefits of BP's compositionality in scalable verification and its applications~\cite{harel_composing_2013}. See also~\cite{qin2020smt} for a similar approach using the BIP modeling framework. However, previous work on BP verification primarily concentrated on explicit model checking. As systems grow in size and complexity, the limitations of explicit model checking become apparent, necessitating the adoption of symbolic verification techniques as an imperative alternative. 

BPpy's symbolic model checker takes a unique approach by utilizing the inherent compositionality of behavioral programs. It independently explores the state space of each b-thread and analyzes the product space symbolically, avoiding the explicit enumeration of all states. More specifically, when applying symbolic model checking in BPpy, b-programs are automatically translated to an SMV model~\cite{mcmillan_smv_1993} and verified symbolically using PyNuSMV~\cite{busard_pynusmv_2013}. The translation maps each b-thread to an SMV module, which reflects the space explored using Depth First Search (DFS). For instance, consider the module presented in \autoref{lst:add-hot-smv}, which represents the \li{add_hot} b-thread in \autoref{lst:hot-cold-intro}. Within each b-thread module, there exists a local variable called \li{state}, capturing the current state of the b-thread. Additionally, each module incorporates local Boolean variables for each event that the associated b-thread may either request or block. For example, since the \li{add_hot} b-thread requests the \li{HOT} event, its corresponding module contains the Boolean variable \li{HOT_requested}, which dynamically changes based on the current \li{state}.

\begin{lstlisting}[
keywords={MODULE,VAR,INIT,DEFINE,TRANS, next, FALSE, TRUE, case, esac,ASSIGN},
label={lst:add-hot-smv},
caption={The \li{add_hot} b-thread translation to SMV.},
float=htbp
]
MODULE add_hot(event)
  VAR
    state: 0 .. 4;
    HOT_requested: boolean;
  INIT
    state = 0
  ASSIGN
    HOT_requested := 
      case
        state = 2 | state = 1 | state = 0 : TRUE;
        state = 3: FALSE;
        TRUE: FALSE;
      esac;
    next(state) := 
      case
        next(event) = HOT : state + 1;
        TRUE: state;
      esac;
\end{lstlisting}

\autoref{lst:main-smv} depicts the main module of the translated SMV model derived from  HOT/COLD b-program discussed in \autoref{sec:bppy}. In the translated model, this main module functions as the event arbiter for the b-program and implements BP semantics. It initiates an enumerated variable \li{event}, representing the currently selected event. The \li{event} variable can take any of the events requested across all b-threads, in addition to two auxiliary events, \li{BPROGRAM_START} and \li{BPROGRAM_DONE}, marking the program's start and end of execution, respectively. 
Further, the main module activates all the b-threads of the program as module instances. It tracks each b-thread instance, requested and blocked variables, by using the \texttt{DEFINE} operator.
To capture the b-program's dynamics, the system defines a transition relation that allows the next \li{event} value to be any currently enabled event. If no such event exists, the b-program terminates, and the system transitions to a sink state where \li{event = BPROGRAM_DONE}. This setting facilitates the detection of possible violations related to the termination of the b-program, such as deadlocks or early termination scenarios.

\begin{lstlisting}[
keywords={MODULE,VAR,INIT,DEFINE,TRANS, next, FALSE, TRUE, case, esac,ASSIGN},
label={lst:main-smv},
caption={The translated main module of the HOT/COLD b-program discussed in \autoref{sec:bppy}.},
float=htbp
]
MODULE main
  VAR
   event:{BPROGRAM_START,BPROGRAM_DONE,HOT,COLD};
   bt0: add_hot(event);
   bt1: add_cold(event);
   bt2: control(event);
  INIT
    event = BPROGRAM_START
  DEFINE
    HOT_requested := bt0.HOT_requested;
    HOT_blocked := bt2.HOT_blocked;
    COLD_requested := bt1.COLD_requested;
    COLD_blocked := bt2.COLD_blocked;
    HOT_enabled := HOT_requested & !HOT_blocked;
    COLD_enabled:=COLD_requested & !COLD_blocked;
  TRANS
    next(event) != BPROGRAM_START & (!HOT_enabled -> next(event) != HOT) & (!COLD_enabled -> next(event) != COLD) & (HOT_enabled | COLD_enabled -> next(event) != BPROGRAM_DONE) & (event = BPROGRAM_DONE -> next(event) = BPROGRAM_DONE)
\end{lstlisting}
We now turn to a performance evaluation for the aforementioned approach. All programs and tools used for evaluation in this study are available in the supplementary material. The evaluation involved several b-programs, including 1) The HOT/COLD example presented in \autoref{sec:bppy} with an increasing number of portions (N) and cold b-threads (M); 2) The Dining Philosophers b-program presented in~\cite{elyasaf_what_2023}; and 3) The Tic-Tac-Toe game presented in~\cite{elyasaf_what_2023}. For this evaluation, all the specifications we considered hold (i.e., no violations are discovered through model checking) to examine how quickly the various approaches can traverse the entire search space. We compared the results against similar b-programs verified using BPjs~\cite{bar-sinai_verification_2021}, a Java-based tool for running and analyzing behavioral programs written in JavaScript, which is the current best practice in BP verification. The symbolic verifier in BPpy can function in two modes: Binary Decision Diagrams (BDD) and SAT-based Bounded Model Checking (BMC), where the model is unrolled for a fixed number of steps and is checked for violations that can occur within that number of steps or fewer. We compared these modes to the BPjs verifier in unbounded and bounded modes, respectively. The results are available in \autoref{tab:model-checking-results}.

\begin{table}[hbt]
    \centering
    \scriptsize
    \setlength{\tabcolsep}{0.4pt} 
\begin{threeparttable}
\begin{tabular}{c|c|c|c||c|c|c|c||c|c|c|c}
\multicolumn{4}{c}{} &  \multicolumn{4}{c}{\bfseries Unbounded	} &   \multicolumn{4}{c}{\bfseries Bounded	}\\
    \multicolumn{4}{c}{} &  \multicolumn{2}{c}{\bfseries time$^3$	} &   \multicolumn{2}{c}{\bfseries memory$^4$	} &  \multicolumn{2}{c}{\bfseries time$^3$	} &   \multicolumn{2}{c}{\bfseries memory$^4$	}\\
     &	 N,M&	 $|E|$$^1$&	 $|S|$$^2$&   BDD &  BPjs &    BDD &  BPjs &   BMC &  BPjs &    BMC &  BPjs \\
        \hline
        \multirow{9}{*}{\begin{tabular}[c]{@{}c@{}}HOT/\\COLD\end{tabular}} & 30,1 & 2 & 121 & \textbf{2.9} & 5.1 & \textbf{0.08} & 0.24 & \textbf{3.8} & 5.2 & \textbf{0.09} & 0.23 \\

        & 60,1 & 2 & 122 & \textbf{3.5} & 5.3 & \textbf{0.08} & 0.29 & \textbf{5.0} & 5.3 & \textbf{0.13} & 0.29 \\

        & 90,1 & 2 & 182 & \textbf{4.2} & 5.6 & \textbf{0.08} & 0.34 & 7.9 & \textbf{5.4} & \textbf{0.19} & 0.36 \\

        & 30,2 & 3 & 1022 & \textbf{3.5} & 10.0 & \textbf{0.08} & 0.78 & \textbf{5.2} & 8.4 & \textbf{0.10} & 1.03 \\

        & 60,2 & 3 & 3842 & \textbf{4.4} & 50.9 & \textbf{0.09} & 1.21 & \textbf{18.3} & 22.6 & \textbf{0.18} & 1.64 \\

        & 90,2 & 3 & 8462 & \textbf{5.3} & 215.6 & \textbf{0.11} & 1.32 & \textbf{99.1} & 102.0 & \textbf{0.37} & 1.65 \\

        & 30,3 & 4 & 11437 & \textbf{3.9} & 398.8 & \textbf{0.08} & 1.19 & \textbf{5.4} & 223.9 & \textbf{0.10} & 1.61 \\

        & 60,3 & 4 & 158901 & \textbf{5.1} & \emph{t.o.} & \textbf{0.11} & \emph{t.o.} & \textbf{22.7} & \emph{t.o.} & \textbf{0.20} & \emph{t.o.} \\

        & 90,3 & 4 & 519151 & \textbf{7.4} & \emph{t.o.} & \textbf{0.13} & \emph{t.o.} & \textbf{121.5} & \emph{t.o.} & \textbf{0.46} & \emph{t.o.} \\
        \hline
        \multirow{3}{*}{\begin{tabular}[c]{@{}c@{}}Dining\\Phil.\end{tabular}} & 3,- & 12 & 106 & \textbf{4.6} & 6.1 & \textbf{0.12} & 0.29 & 5.5 & \textbf{4.5} & \textbf{0.08} & 0.25 \\

        & 6,- & 36 & 30862 & \emph{t.o.} & \textbf{56.8} & \emph{t.o.} & \textbf{0.97} & 8.0 & \textbf{5.0} & \textbf{0.09} & 0.30 \\

          & 9,- & 54 & 3299501 & \emph{t.o.} & \emph{o.m.} & \emph{t.o.} & \emph{o.m.} & 11.2 & \textbf{5.8} & \textbf{0.10} & 0.36 \\
        \hline
        TTT & 3,- & 21 & 69502 & \emph{t.o.} & \textbf{1852.1} & \emph{t.o.} & \textbf{10.01} & \textbf{26.1} & 750.6 & \textbf{0.15} & 7.98 \\

        \hline
    \end{tabular}
\begin{scriptsize}
    $^1$ the number of program events, $^2$ the number of program states, $^3$ in seconds, $^4$ in GB, \emph{o.m.} out of memory (16GB), \emph{t.o.} timeout (60 minutes)

\end{scriptsize}

\end{threeparttable}
     \caption{The average time and memory  (over 10 repetitions) required to verify b-programs in BPpy and BPjs. The Binary Decision Diagrams (BDD) and SAT-based Bounded Model Checking (BMC) modes implemented in BPpy are compared against the BPjs verifier in unbounded and bounded modes, respectively.
     }
     \label{tab:model-checking-results}
\end{table}

In terms of memory consumption, both symbolic model checkers outperformed the explicit verifier, especially the BDD-based verifier, exhibiting improved efficiency. This result aligns with expectations, given that the program's state space is symbolically represented through BDDs or logical formulas rather than being explicitly enumerated. Regarding verification time, we observed that the BDD-based model checker excelled in the examples where the total number of events remained relatively small. However, as the number of events increased, the verification time for the BDD-based model checker experienced a significant surge. Conversely, when considering the bounded option, BMC demonstrated greater resilience to a growing number of events in the b-program, positioning itself as a more robust approach.

The translation of B-programs into SMV models, utilizing their inherent compositionality, shows promise for achieving scalability and efficiency in verification. Our future research will focus on enhancing the translation process to improve the verification of programs with a larger number of events. Also, we aim to extend the symbolic model checking support to various execution mechanisms and protocols, including the one outlined in Section \ref{sec:smt}.

\vspace{-10pt}
\section{BP \texorpdfstring{$\Leftrightarrow$}{<->} Probabilistic Modeling}
\label{sec:probabilistic-modeling}
\vspace{-3pt}
This section discusses BP as a modular language for modeling and analyzing probabilistic systems, enhancing its semantics for compositional modeling of probabilistic and non-deterministic behaviors. Traditionally, randomness in BP is introduced through a random function for event selection, with the \li{sync} node's standard event selection protocol being an example. However, this is limited as it only allows uniform distributions. Since we want to allow non-uniform distribution, we take the alternative approach of allowing a random behavior of the b-threads. We want to define randomness in a way that allows both execution (sampling) and model checking. This poses a non-trivial challenge because, during model checking, the analysis needs to consider all possible paths with their probabilities. Thus, we introduce the \li{choice} idiom, illustrated in \autoref{lst:coin-flip}, where a dictionary maps values to probabilities for defining categorical distributions in b-threads. These distributions can be sampled during the execution of a b-program or transformed for analysis.

\begin{lstlisting}[
label={lst:coin-flip},
caption={Uneven coin flip using the choice statement.},
float=htbp
]
def coin_flip():
  side = yield choice({"heads":0.4,"tails":0.6})
  yield sync(request=BEvent(side))
\end{lstlisting}

The proposed BP model combines non-determinism and probability for different modeling purposes. Non-determinism is handled through \li{sync} points with multiple events, while probabilities are specified using \li{choice}. Non-determinism arises in systems when there is a decision to make. Probabilities are used to model uncontrolled random events. Thus, during analysis, non-determinism is typically addressed by analyzing contingencies, while probabilities are computed for each path. 

Executing programs with \li{choice} statements involves random selection based on a specified distribution, facilitating naive sampling. Model analysis is done by translation into the PRISM language~\cite{kwiatkowska_prism_2011}, a tool for analyzing probabilistic systems with a module-based approach. The translation process, largely similar to the one discussed for SMV in \autoref{sec:model-checking}, has been automated and integrated into BPpy. Supplementary material includes technical details about this process, code, and documentation. We utilize the \li{mdp} model type in PRISM to analyze systems that involve a combination of non-determinism and probabilistic behavior.                     

We compared sampling and formal analysis to evaluate translation performance. Experiments timed the analysis of parametrized versions of three models. Sampling involved running models repeatedly and monitoring outcomes. Formal verification translated the model into PRISM format and verified it using Storm probabilistic model checker~\cite{hensel_probabilistic_2022}. All programs and tools used for this evaluation can be found in the supplementary material.

The first model, the classic Monty Hall problem, which is infamous for its counter-intuitive solution, is described as follows: ``A game host hides a prize behind one of three doors. The contestant guesses which door has the prize. The host then opens one door with no prize. The contestant can stick with their choice or switch.'' Our evaluation parameterized the total number of doors, prizes, and doors ruled out, as proposed in~\cite{depuydt_higher_2012}. The model presented in \autoref{lst:monty-hall} consists of three b-threads. The first b-thread, named \li{hide_prizes}, determines where to hide prizes, requests the hiding events, blocks the opening of the doors with hidden prizes until the opening phase is complete, and waits for the contestant to open a door. It checks if the chosen door contains a prize and accordingly announces a \li{win} or \li{lose} event. We added the \li{repeat}, \li{replace}, and \li{sorted} parameters to the \li{choice} statement to ease the use of the idiom in cases where repeated sampling is required. The \li{make_a_guess} b-thread waits for the end of the hiding phase, guesses a door, and then blocks its opening. The \li{open_doors} b-thread waits for the guess and then requests to open doors while blocking the already opened doors. Then, it requests the event that marks the end of the host door opening phase and the event for the door the contestant eventually opens.

\begin{lstlisting}[
label={lst:monty-hall},
caption={A BPpy model of the Monty Hall problem.},
float=htbp
]
@thread
def hide_prizes(doors, prizes_num):
  prizes = yield choice({i: 1/len(doors) for i in doors},
       repeat=prizes_num, replace=False, sorted=True)
  for hide in prizes:
    yield sync(request=BEvent(f"hide{hide}"))
  yield sync(request=BEvent("done_hiding"))
  dont_open = [BEvent(f"open{d}") for d in prizes]
  yield sync(block=dont_open,waitFor=BEvent("done_opening")
  door = yield sync(waitFor=all_open)
  yield sync(request=BEvent("win" if int(door.name[4:]) in prizes else "lose"))

@thread
def make_a_guess():
  yield sync(waitFor=BEvent("done_hiding"))
  yield sync(request=BEvent(f"guess{0}"))
  yield sync(block=BEvent(f"open{0}"))

@thread
def open_doors(doors, doors_opened_num):
  yield sync(waitFor=[BEvent(f"guess{d}") for d in doors])
  blocked = []
  for _ in range(doors_opened_num):
    e = yield sync(request=all_open, block=blocked)
    blocked += [e]
  yield sync(request=BEvent("done_opening"))
  yield sync(request=all_open, block=blocke)
\end{lstlisting}

The experiment included generating instances of the b-program from \autoref{lst:monty-hall} for all parameter combinations up to 10 doors. For sampling, we executed the program 10,000 times to track the \li{win} event occurrence and used PRISM and Storm for model checking to calculate the probability of reaching the \li{win} state. The analysis time is the entire duration of Storm.

\begin{figure*}[hbt]
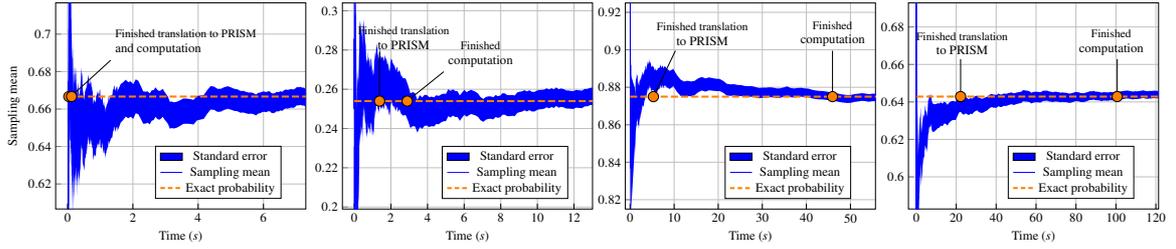

    \centering
    \captionsetup[subfigure]{justification=centering}
    \begin{subfigure}[t]{0.25\textwidth}
       \input{3d1p1o}
       \vspace*{-5mm}
        \caption{$d=3,p=1,o=1$}
        \label{fig:monty_runtime_311}
    \end{subfigure}
    \begin{subfigure}[t]{0.23\textwidth}
        \input{9d2p1o}
       \vspace*{-5mm}
        \caption{$d=9,p=2,o=1$}
        \label{fig:monty_runtime_921}
    \end{subfigure}
    \begin{subfigure}[t]{0.23\textwidth}
        \input{8d4p3o}
       \vspace*{-5mm}
        \caption{$d=8,p=4,0=3$}
        \label{fig:monty_runtime_843}
    \end{subfigure}
    \begin{subfigure}[t]{0.23\textwidth}
        \input{10d5p2o}
       \vspace*{-5mm}
        \caption{$d=10,p=5,0=2$}
        \label{fig:monty_runtime_1052}
    \end{subfigure}
    \caption{A comparison of the sampling-based analysis and the exact analysis for the Monty Hall model with different values of doors ($d$), prizes ($p$), and doors opened ($o$).\vspace{-5pt}}
    \label{fig:monty_runtime}
\end{figure*}

The results of applying two analysis methods to the Monty Hall model with different parameter values are shown in \autoref{fig:monty_runtime}. The blue line and halo depict the mean and standard error of collected samples over time. The dashed orange line shows the exact computation value, and the orange circles mark the time of translation to PRISM and the subsequent analysis. These findings indicate that exact analysis yields results before sampling variance reaches acceptable error margins in all examined cases.

\autoref{tab:monty-results} provides a breakdown of the runtime for the exact analysis in the proposed approach for the Monty Hall model with 10 doors, distinguishing between PRISM model construction and Storm analysis. An expected correlation exists between the time spent on model translation and overall computation time. The primary source of complexity appears to be the number of doors opened, as the modeling process times out entirely after 6 doors. This is likely due to numerous different paths that do not merge, unlike the situation with the number of prizes.

\begin{table}[hbt]
\begin{center}
\footnotesize
\setlength{\tabcolsep}{1pt}
\centering
\scriptsize
\begin{threeparttable}
    \begin{tabular}{c|c|c|c|c|c|c}

\multicolumn{1}{c}{}&  \multicolumn{6}{c}{\textbf{Opened}}\\
\textbf{Prizes}& \textbf{1} & \textbf{2} & \textbf{3} & \textbf{4} & \textbf{5} & \textbf{6}        \\ \hline
\textbf{1}                                                        & 0.6/0.1   & 1.1/0.5   & 9.4/8.4    & 114.7/536.3  & 4120.6/$t.o.$ & $t.o.$/$t.o.$ \\ \hline
\textbf{2}                     & 2.2/1.3   & 4.1/3.8    & 8.7/46.0   & 112.0/1401.4 & 3798.0/$t.o.$ & $t.o.$/$t.o.$ \\ \hline
\textbf{3}                     & 6.5/5.9   & 7.6/18.3  & 13.2/136.3  & 122.1/$t.o.$     & 4376.5/$t.o.$ & $t.o.$/$t.o.$ \\ \hline
\textbf{4}                     & 13.9/19.  & 15.5/52.8 & 21.0/252.7  & 123.3/$t.o.$      & 3874.2/$t.o.$ &          \\ \hline
\textbf{5}                     & 21.1/29.4 & 22.2/64.2 & 28.6/197.2  & 135.5/$t.o.$     &              &          \\ \hline
\textbf{6}                     & 17.8/19.9 & 18.2/34.5 & 24.7/77.5   &                 &              &          \\ \hline
\textbf{7}                     & 10.1/6.4  & 10.9/8.4  &             &                 &              &          \\ \hline
\textbf{8}                     & 8.3/1.4   &           &             &                 &              &   \\
\end{tabular}
\begin{scriptsize}
\emph{t.o.} timeout (two hours)
\end{scriptsize}
\end{threeparttable}
\end{center}
    \caption{
    Runtime in seconds for exact analysis with various parameter combinations for the Monty Hall model with 10 doors. The table presents PRISM model construction times on the left and Storm analysis times on the right.}
    \label{tab:monty-results}
\end{table}

\begin{figure}
    \centering
    \begin{tikzpicture}[scale=0.35]

\node at (2,10) {\scriptsize 1,0} [sibling distance = 8cm]
	child {node (C) {\scriptsize 2,0} [sibling distance = 4cm]
        child {node {\scriptsize 4,0} [sibling distance = 2cm]
            child {node [blue] {\scriptsize 8,0}}
            child {node [blue] {\scriptsize 8,1}}}
        child {node {\scriptsize 4,1} [sibling distance = 2cm]
            child {node [blue] {\scriptsize 8,2}}
            child {node [blue] {\scriptsize 8,3}}
        }
    }
	child {node (D) {\scriptsize 2,1} [sibling distance = 4cm]
        child {node {\scriptsize 4,2} [sibling distance = 2cm]
            child {node [blue] {\scriptsize 8,4}}
            child {node [blue] {\scriptsize 8,5}}}
        child {node {\scriptsize 4,3} [sibling distance = 2cm]
            child {node (A) [red] {\scriptsize 8,6}}
            child {node (B) [red] {\scriptsize 8,7}}
        }
    };
\draw [dashed, ->] (A) to[bend right=25] (C);
\draw [dashed, ->] (B) to[bend right=45] (D);

\end{tikzpicture}
    \caption{An illustration of the process simulating a six-sided dice using coin flips.}
    \label{fig:dice-tree}
\end{figure}
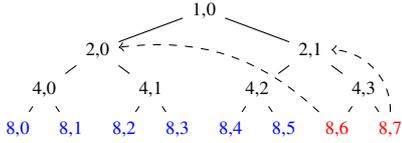

Our second evaluation involves modeling a fair $n$-sided dice using fair coins, following Knuth's algorithm~\cite{knuth1976complexity} featured in PRISM. This algorithm uses rejection sampling to minimize the expected coin tosses by simulating a uniform distribution over powers of two with fair coins. For example, in the case of a six-sided die (see \autoref{fig:dice-tree}), three coin tosses yield 8 possibilities. If the resulting value of the tosses is smaller than six, the algorithm accepts it. Otherwise, the algorithm repeats the process as if starting with the remainder. More complex cases may involve multiple trees for simulation.

The BP model for this algorithm tracks coin flips through b-threads representing nodes in the sequence. Each node waits for its parent. Then, depending on its place in the tree, it either returns the dice value or performs a toss and requests the corresponding next node. This results in a simple model where all b-threads can be generated by a single function shown in \autoref{lst:dice-program}.

\begin{lstlisting}[,
label={lst:dice-program},
caption={B-thread definition of any single node in the BP model of Knuth's algorithm \cite{knuth1976complexity}.},
float=htbp
]
@thread
def node(u, x): # u: layer size, x: index in layer
 while True:
  yield sync(waitFor=BEvent(f"n{u}_{x}"))
  if u < n: # inner node
   flip = yield choice({0:0.5, 1:0.5})
   yield sync(request=BEvent(f"n{u*2}_{2*x+flip}")
  else: # last layer
   if x >= n:
    yield sync(request=BEvent(f"n{u-n}_{x%n}"))
   else:
    yield sync(request=BEvent(f"result_{x}"))
\end{lstlisting}

\begin{figure}[hbt]
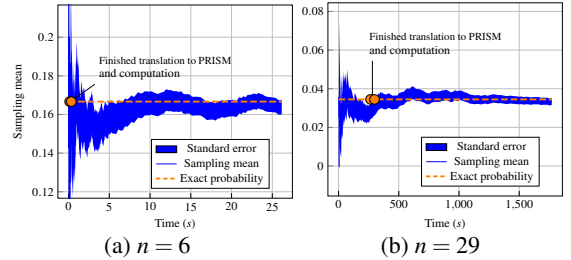

    \centering
    \captionsetup[subfigure]{justification=centering}
        \centering
        \begin{subfigure}[t]{0.23\textwidth}
       \centering
       \input{6}
       \vspace{-6mm}
        \caption{$n=6$}
        \label{fig:dice_runtime_6}
        \end{subfigure}
            \begin{subfigure}[t]{0.23\textwidth}
       \centering
        \input{29}
       \vspace{-6mm}
        \caption{$n=29$}
        \label{fig:dice_runtime_29}
    \end{subfigure}
    \caption{Comparison of the sampling-based and exact analysis for the $n$-sided dice model.}
    \label{fig:dice_runtime}
    
\end{figure}

Results from the dice program analysis in \autoref{tab:dice-results} and \autoref{fig:dice_runtime} align with our observations from smaller Monty Hall versions, where the exact analysis outperforms sampling. One difference from the previous example was that translation, rather than computation, was the more time-consuming part of the exact analysis due to more b-threads and fewer commonly shared events. This example is notable due to the unlimited number of times it may repeat before yielding a result, making it particularly suited for exact analysis, which explores the full possibility space simultaneously. We can observe that in more complex cases like \autoref{fig:dice_runtime_29}, the exact analysis still performs well due to oversampling advantages.

\begin{table}[hbt]
\begin{center}
\footnotesize
\setlength{\tabcolsep}{2pt}
\centering
\scriptsize
\begin{tabular}{c|c|c|c c c|c|c|c}
\textbf{$n$} & \textbf{States}
& \textbf{\begin{tabular}[c]{@{}c@{}}Trans./Comp.\\ time (s)\end{tabular}}
& \textbf{Result} & & \textbf{$n$} & \textbf{States}
& \textbf{\begin{tabular}[c]{@{}c@{}}Trans./Comp.\\ time (s)\end{tabular}}
& \textbf{Result} \\ \cline{1-4}\cline{6-9} 
\textbf{6 } & 64   & 0.129/0.21   & 0.1667 & & \textbf{19} & 1868 & 46.931/6.056 & 0.0526 \\ 
\textbf{7 } & 68   & 0.141/0.053  & 0.1429 & & \textbf{20} & 512  & 3.179/0.418  & 0.05     \\ 
\textbf{8 } & 38   & 0.132/0.053  & 0.125    & & \textbf{21} & 440  & 3.009/0.399  & 0.0476 \\ 
\textbf{9 } & 292  & 1.361/0.214  & 0.1111 & & \textbf{22} & 1200 & 20.060/2.451 & 0.0455 \\ 
\textbf{10} & 216  & 0.821/0.147  & 0.1      & & \textbf{23} & 968  & 12.796/1.522 & 0.0435 \\ 
\textbf{11} & 596  & 5.095/0.635  & 0.0909 & & \textbf{24} & 376  & 1.592/0.233  & 0.0417 \\ 
\textbf{12} & 168  & 0.423/0.097  & 0.0833 & & \textbf{25} & 2744 & 101.133/12.561& 0.04     \\ 
\textbf{13} & 848  & 9.941/1.191  & 0.0769 & & \textbf{26} & 1704 & 39.804/5.134 & 0.0385 \\ 
\textbf{14} & 144  & 0.440/0.097  & 0.0714 & & \textbf{27} & 2668 & 96.446/12.325& 0.0370 \\ 
\textbf{15} & 148  & 0.443/0.099  & 0.0667 & & \textbf{28} & 368  & 1.664/0.239  & 0.0357 \\ 
\textbf{16} & 78   & 0.451/0.097  & 0.0625   & & \textbf{29} & 4448 & 265.348/33.184& 0.03448 \\ 
\textbf{17} & 752  & 8.300/0.999  & 0.0588 & & \textbf{30} & 304  & 1.705/0.237  & 0.03333 \\ 
\textbf{18} & 592  & 5.272/0.663  & 0.0556 & &  &   &    &  \\

\end{tabular}

\end{center}
    \caption{
    Exact analysis of various sizes for a dice. Some require multiple trees which repeat the same structure.}
    \label{tab:dice-results}
\end{table}

Our last case study is a variation of the bit-flip scenario discussed in \autoref{sec:smt}. In this variation, diagonals are initiated randomly, and the actions are taken at each step, where each consecutive row or column can be flipped. The goal here is to find a sequence that results in the board having the same (or as close as possible for odd-size boards) number of bits turned on and off.

The translation part in the exact analysis of the bit-flip model was more challenging than in previous cases, starting at relatively small matrix sizes. This is due to the large number of board configurations in this scenario. Each of the possible $2^{n\times m}$ possibilities adds both an event and a state in each b-thread, resulting in an exponential number of options that must be explored in all of the b-threads during translation. However, when considering all b-threads as a unified b-program during exact analysis, the number of possible outcomes is relatively small.

We conducted a sampling-based analysis by measuring $9,999$ runs or as many as possible within an hour, as seen in \autoref{tab:bitflip-sampling}. Sampling outperformed exact analysis, supporting double the translation's grid size before slowing down. We suspect the difficulties stem from the sync events being a product of relatively costly computations. \autoref{sec:smt} and \autoref{sec:four-way-integration} present an alternative analysis approach to the discrete event selection mechanism used here by integrating SMT solvers, which aims to address this issue.

\begin{table}[hbt]
\begin{center}
\footnotesize
\setlength{\tabcolsep}{2pt}
\centering
\scriptsize
\begin{tabular}{c|c|c|c|c c c|c|c|c|c}
\textbf{Dim}  & \textbf{Mean} & \textbf{SEM} & \textbf{Sam.} & \textbf{Time} && \textbf{Dim}  & \textbf{Mean} & \textbf{SEM} & \textbf{Sam.} & \textbf{Time} \\ \cline{1-5} \cline{7-11}
\textbf{2x2}  & 0.263         & 0.004        & 9999             & 0.002             && \textbf{3x3}  & 0.441         & 0.005        & 9999             & 0.023 \\ 
\textbf{2x3}  & 0.264         & 0.004        & 9999             & 0.004             && \textbf{3x4} & 0.178         & 0.004        & 9999             & 0.295 \\
\textbf{2x4}  & 0.309         & 0.005        & 9999             & 0.008             && \textbf{3x5} & 0.329         & 0.014        & 1062             & 3.39 \\ 
\textbf{2x5}  & 0.309         & 0.005        & 9999             & 0.041             && \textbf{3x6} & 0.197         & 0.052        & 61               & 59.934 \\ 
\textbf{2x6}  & 0.241         & 0.004        & 9999             & 0.237             && \textbf{3x7} & 0.4           & 0.25         & 5                & 885.183 \\ 
\textbf{2x7}  & 0.124         & 0.007        & 2476             & 1.454             && \textbf{4x4}  & 0.209         & 0.024        & 292              & 12.34 \\ 
\textbf{2x8}  & 0.081         & 0.013        & 456              & 7.903             && \textbf{4x5}  & 0.143         & 0.167        & 7                & 554.663 \\ 
\textbf{2x9}  & 0.13          & 0.051        & 46               & 79.496            &&   &          &        &                 &  \\ 
   
\end{tabular}
\end{center}
    \caption{
    Sampling results of the discrete bit-flip model with: average, standard measurement error, number of samples (Samp.), and average time per sample in seconds (Time).
    }
    \label{tab:bitflip-sampling}
\end{table}

The presented examples cover a wide range of parameter values and diverse model types, spanning from models with few states and minimal transitions, such as the dice problem, to highly interconnected ones in bit-flip. Our methods show promise in different problem domains, especially in smaller systems. 

Modeling probabilistic systems with both non-deterministic and probabilistic behavior has been addressed in various ways previously. Some models focus on representing systems with probabilistic transitions~\cite{PUTERMAN1990331,stoelinga_introduction_2004}, while others are designed to represent systems with non-deterministic choices~\cite{rabinandscott,savitch1970relationships}. For systems exhibiting both non-deterministic and probabilistic behavior, models such as PRISM~\cite{kwiatkowska_prism_2011} are commonly used. These models allow for a rich representation of system dynamics, enabling the analysis and verification of system properties. The advantages of modeling randomness in BP have not been explored before. The only reference to randomness we know of is in event selection strategies~\cite{harel_programming_2010}.

\vspace{-10pt}
\section{BP \texorpdfstring{$\Leftrightarrow$}{<->} DRL}
\label{sec:drl}
\vspace{-3pt}
This section explores the interplay between BP and deep reinforcement learning (DRL), focusing on how it can enhance the alignment of requirements with BP modules for more effective execution. We illustrate how DRL's advanced data processing and interpretation abilities can aid in designing complex systems in BP. Subsequently, we discuss how BP's structured way of encoding rich behavioral specifications can enhance DRL-driven solutions. This is demonstrated using BPpy's integration with Gymnasium~\cite{towers_gymnasium_2023}, a widely recognized API standard for reinforcement learning (RL) environments.

The combination of RL and BP was first introduced by~\cite{eitan_adaptive_2011}, which enhanced the semantics of live sequence charts (LSC), a visual language for BP, by incorporating reinforcements and applying learning algorithms. In recent years, this combination has gained growing interest, with several studies conducted with the assistance of BPpy. These studies explored different ways this learning mechanism can be combined with BP's execution~\cite{elyasaf_using_2019,yaacov_keeping_2024}.

To illustrate how DRL can enhance the efficiency of execution mechanisms in BP, we begin with an evaluation of the \emph{Blueberry Pancake Maker} example presented in~\cite{bar-sinai_extending_2020}. In this example, a pancake batter is made from dry and wet mixtures using a mixer controlled by a b-program. The first two b-threads in \autoref{fig:pancake-a} prepare batter for $n$ pancakes by adding $n$ mixtures without specifying the order, which can lead to improper batter thickness. To prevent damage to the mixer, we need to control the order of mixture additions. The \li{thickness_meter} b-thread monitors thickness changes, while the \li{range_arbiter} b-thread ensures the thickness stays within $[-b, b]$. Next, we add blueberries to the mix. To prevent the blueberries from bursting during the mixing process, 75\% of the total mixture of batter should be added, and the batter itself should be relatively thin. The b-threads \li{enough_batter} and \li{batter_thin_enough} block the blueberry addition event until these conditions are met. Depending on the sequence in which the dry and wet mixtures are added, scenarios may arise where the \li{AddBlueberries} event, requested by the \li{blueberries} b-thread, remains blocked by \li{batter_thin_enough} throughout the program's execution. Such a situation would inevitably violate the system requirements, as it would result in a pancake mixture with no blueberries.

\begin{lstlisting}[,
label={fig:pancake-a},
caption={The \emph{Blueberry Pancake Maker} b-program~\cite{bar-sinai_extending_2020}. },
float=htbp
]
@thread
def add_dry_mixture(n):
 for i in range(n):
  yield sync(request=DryMixture())

@thread
def add_wet_mixture(n):
 for i in range(n):
  yield sync(request=WetMixture())

@thread
def thickness_meter():
 while True:
  e = yield sync(waitFor=mixture_add)
  if e == DryMixture() :
   yield sync(request=ThicknessUp(),block=mixture_add)
  else:
   yield sync(request=ThicknessDown,block=mixture_add)
   
@thread
def range_arbiter(b):
 thickness = 0
 while True:
  e = yield sync(waitFor=any_thick)
  thickness += 1 if e == ThicknessUp() else -1
  if abs(thickness) >= b:
   block_e = DryMixture() if thickness>0 else WetMixture() 
   yield sync(block=block_e, waitFor=mixture_add)
  else:
   yield sync(waitFor=mixture_add)

@thread
def blueberries():
 yield sync(request= AddBlueberries())

@thread
def enough_batter(n):
 for j in range(int((n * 3) / 2)):
  yield sync(waitFor=mixture_add,block=AddBlueberries())

@thread
def batter_thin_enough(n):
 thickness = 0
 while True:
  if thickness >= 0:
   e=yield sync(waitFor=any_thick,block= AddBlueberries())
  else:
   e=yield sync(waitFor=any_thick)
  thickness += 1 if e.name=="ThicknessUp" else -1
\end{lstlisting}

To address the risk of missing blueberries, one option is to define a specific sequence for their addition. However, this becomes progressively more complicated with larger systems and various requirements. For instance, if dry and wet mixtures are not consistently ready for addition, determining a strict sequence becomes more challenging and prone to errors. Thus, our aim is to design flexible b-programs to allow for all valid behaviors.

To address the complexity in specifying that blueberries must eventually be added, we used the integration between BPpy and Gymnasium introduced in~\cite{yaacov_bppy_2023}. This integration encapsulates a b-program as an RL environment, allowing developers to delineate the program's target objectives or optimization criteria by incorporating the \li{localReward} parameter. This parameter can be integrated into any yield statement, complementing traditional requirements and facilitating learning through RL algorithms. Embedding these reward-based criteria within BPpy streamlines the modeling of intricate systems, fostering more intuitive and efficient development practices. \autoref{lst:pancake4} demonstrates how rewards are added to the \li{blueberries} b-thread, reflecting the system's preference for eventually adding blueberries.

\begin{lstlisting}[,
label={lst:pancake4},
caption={The updated \li{blueberries} b-thread which the \li{localReward} parameter. },
float=htbp
]
@thread
def blueberries():
 yield sync(request= AddBlueberries(), localReward=-0.0001)
 yield sync(waitFor=All(), localReward=1)
\end{lstlisting}

Our evaluation focused on the task of finding a deterministic event selection mechanism to generate a single execution trace adhering to system requirements. We compared two mechanisms. The first uses a DRL algorithm, wherein the state space is represented by local variables of the b-threads, serving as input to the agent's neural network. The learning algorithm was aborted upon successful task completion, producing a correct sequence of events in its predictions. Specifically, we employed the Maskable PPO algorithm~\cite{huang_closer_2022} implemented in the Stable Baselines3 package~\cite{hill_stable_2018}, with a standard multilayer perception (MLP) network with two hidden layers of size 64. The second mechanism we evaluated uses an explicit program synthesis approach, in which a valid execution trace is found by exploring the program's state space. This is considered the current standard approach, as it has been discussed previously, and is supported by existing tools.


The runtime and memory usage of the two mechanisms for finding a single execution trace are presented in \autoref{tab:drl-results-1}. Tests were conducted with $n \in \{200,300,400,500\}$ and $b \in \{25,50,75,100\}$. As expected, the runtime and memory of the synthesis approach increase with the problem size, especially $n$. In contrast, the DRL approach's runtime and memory remain relatively stable. This can be attributed to the fact that increasing $n$ and $b$ doesn't change the number of program variables, keeping the neural network complexity constant. This demonstrates that DRL can be particularly beneficial when system complexity is due to variable value range rather than quantity.

\begin{table}[!ht]
    \centering
    \scriptsize
    \setlength{\tabcolsep}{2pt} 
\begin{threeparttable}
\begin{tabular}{c|c|c|c|c|c c c|c|c|c|c|c}
\multicolumn{2}{c}{} &  \multicolumn{2}{c}{\bfseries Time$^1$	} &   \multicolumn{2}{c}{\bfseries Memory$^2$		} & \multirow{8}{*}{ }& \multicolumn{2}{c}{} &  \multicolumn{2}{c}{\bfseries Time$^1$	} &   \multicolumn{2}{c}{\bfseries Memory$^2$		}\\
    \bfseries $n$ &	\bfseries $b$ &  \bfseries DRL & \bfseries Syn. & \bfseries DRL & \bfseries Syn. & & \bfseries $n$ &	\bfseries $b$ &  \bfseries DRL & \bfseries Syn. & \bfseries DRL & \bfseries Syn. \\
        \cline{1-6}\cline{8-13} 
        \multirow{4}{*}{200} & 25  & 17   & 123  & 1.79   & 0.06 & & \multirow{4}{*}{400} & 25  & 16   & 555  & 1.79   & 0.11 \\
         & 50  & 19   & 142  & 1.79   & 0.06  & &  & 50  & 18   & 569  & 1.79   & 0.12 \\
         & 75  & 19   & 149  & 1.79   & 0.06 & &  & 75  & 16   & 558  & 1.79   & 0.12 \\
         & 100 & 18   & 163  & 1.79   & 0.06 & &  & 100 & 17   & 591  & 1.79   & 0.12  \\
        \cline{1-6}\cline{8-13}
        \multirow{4}{*}{300} & 25  & 16   & 333  & 1.79   & 0.08 & & \multirow{4}{*}{500} & 25  & 15   & 894  & 1.79   & 0.15 \\
         & 50  & 20   & 312  & 1.79   & 0.08 & & & 50  & 18   & 927  & 1.79   & 0.15 \\
         & 75  & 19   & 339  & 1.79   & 0.09 & &  & 75  & 19   & 873  & 1.79   & 0.15 \\
         & 100 & 20   & 333  & 1.79   & 0.09 & &  & 100 & 16   & 903  & 1.79   & 0.16 \\
        \cline{1-6}\cline{8-13}

    \end{tabular}
\begin{scriptsize}
    $^1$ in seconds, $^2$ in GB
\end{scriptsize}
\end{threeparttable}
     \caption{ Comparison of average (over 10 repetitions) of runtime and memory of the DRL and program synthesis approaches for finding a single valid execution trace.}
     \label{tab:drl-results-1}
\end{table}

We also evaluated the \emph{Cinderella-Stepmother problem} from \autoref{sec:smt}. In this variation, the DRL mechanism needs to effectively handle a counter-strategy of the stepmother described in~\cite{shevrin2020spectra}, where in each round, the stepmother maintains two buckets, which Cinderella cannot empty at once, evenly filled. Once the stepmother reaches a round in which one of the buckets can be entirely filled, it does so and wins. The results of this example experiment are provided as an appendix in the supplementary material. The evaluation yielded results similar to the pancake maker example. 
The appendix also contains early results for the task of finding a non-deterministic event selection mechanism that adheres to system requirements for the two examples. The results show that in both cases, the algorithms converged to accurate strategies after a short training time.

The evolving DRL field presents challenges in the safety, robustness, and interpretability of policies. Previous research suggests using BP to encode expertise for enhancing training and system reliability~\cite{yerushalmi_enhancing_2023,ashrov_enhancing_2023}. We note that BPpy supports knowledge encoding by modifying rewards, features, and actions before relaying them to the learning algorithm or environment, enhancing system safety and ease of training.

\vspace{-10pt}
\section{BP \texorpdfstring{$\Leftrightarrow$}{<->} (DRL+Probabilities+SMT)}
\label{sec:four-way-integration}
\vspace{-3pt}

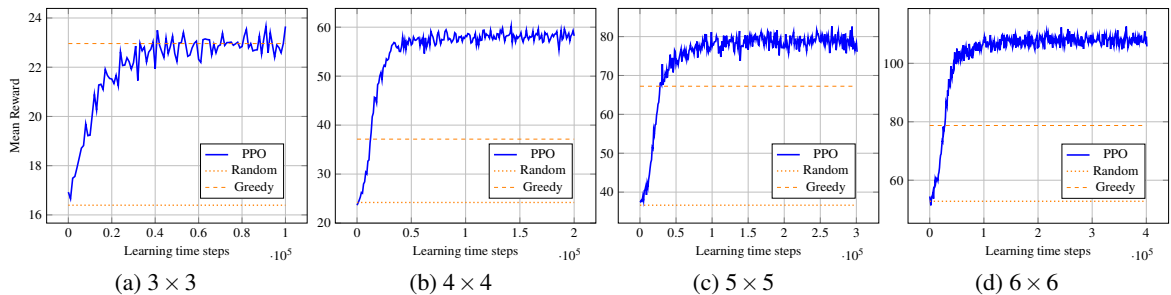
\begin{figure*}[h!]
    \centering
    \captionsetup[subfigure]{justification=centering}
    \begin{subfigure}[t]{0.25\textwidth}
        \centering
        \begin{tikzpicture}[scale=0.5]
        \begin{axis}[
            xlabel={Learning time steps},
            ylabel={Mean Reward},
            grid=major,
            legend style={at={(0.75,0.37)}, anchor=north},
        ]
        \addplot [color=blue, very thick] table [x=timesteps, y=mean_reward, col sep=comma] {bit_flip_drl_3.csv};
        \addlegendentry{PPO}

        \addplot [color=orange, dotted, line width=1pt ] table [x=timesteps, y=mean_reward, col sep=comma] {bit_flip_random_3.csv};
        \addlegendentry{Random}
        \addplot [color=orange, dashed] table [x=timesteps, y=mean_reward, col sep=comma] {bit_flip_greedy_3.csv};
        \addlegendentry{Greedy}
        
        \end{axis}
        \end{tikzpicture}
        \vspace*{-5mm}
        \caption{$3 \times 3$\vspace{.15cm}}
    \end{subfigure}
    \begin{subfigure}[t]{0.23\textwidth}
        \centering
        \begin{tikzpicture}[scale=0.5]
        \begin{axis}[
            xlabel={Learning time steps},
            grid=major,
            legend style={at={(0.75,0.37)}, anchor=north},
        ]
        \addplot [color=blue, very thick] table [x=timesteps, y=mean_reward, col sep=comma] {bit_flip_drl_4.csv};
        \addlegendentry{PPO}

        \addplot [color=orange, dotted, line width=1pt ] table [x=timesteps, y=mean_reward, col sep=comma] {bit_flip_random_4.csv};
        \addlegendentry{Random}
        \addplot [color=orange, dashed] table [x=timesteps, y=mean_reward, col sep=comma] {bit_flip_greedy_4.csv};
        \addlegendentry{Greedy}
        \end{axis}
        \end{tikzpicture}
        \vspace*{-5mm}
        \caption{$4 \times 4$\vspace{.15cm}}
    \end{subfigure}
    \begin{subfigure}[t]{0.23\textwidth}
        \centering
        \begin{tikzpicture}[scale=0.5]
        \begin{axis}[
            xlabel={Learning time steps},
            grid=major,
            legend style={at={(0.75,0.37)}, anchor=north},
        ]
        \addplot [color=blue, very thick] table [x=timesteps, y=mean_reward, col sep=comma] {bit_flip_drl_5.csv};
        \addlegendentry{PPO}

        \addplot [color=orange, dotted, line width=1pt ] table [x=timesteps, y=mean_reward, col sep=comma] {bit_flip_random_5.csv};
        \addlegendentry{Random}
        \addplot [color=orange, dashed] table [x=timesteps, y=mean_reward, col sep=comma] {bit_flip_greedy_5.csv};
        \addlegendentry{Greedy}
        \end{axis}
        \end{tikzpicture}
        \vspace*{-5mm}
        \caption{$5 \times 5$\vspace{.15cm}}
    \end{subfigure}
    \begin{subfigure}[t]{0.23\textwidth}
        \centering
        \begin{tikzpicture}[scale=0.5]
        \begin{axis}[
            xlabel={Learning time steps},
            grid=major,
            legend style={at={(0.75,0.37)}, anchor=north},
        ]
        \addplot [color=blue, very thick] table [x=timesteps, y=mean_reward, col sep=comma] {bit_flip_drl_6.csv};
        \addlegendentry{PPO}

        \addplot [color=orange, dotted, line width=1pt ] table [x=timesteps, y=mean_reward, col sep=comma] {bit_flip_random_6.csv};
        \addlegendentry{Random}
        \addplot [color=orange, dashed] table [x=timesteps, y=mean_reward, col sep=comma] {bit_flip_greedy_6.csv};
        \addlegendentry{Greedy}
        \end{axis}
        \end{tikzpicture}
        \vspace*{-5mm}
        \caption{$6 \times 6$\vspace{.15cm}}
    \end{subfigure}
    \caption{Mean reward of the PPO algorithm for the bit-flip two-player game on square matrices of sizes $3 \times 3$, $4 \times 4$, $5 \times 5$, and $6 \times 6$. The results are compared with random and greedy baseline strategies. 
    }
    \label{fig:bit-flip-drl}
\end{figure*}

In the previous sections, we saw four integrations of BP with other mechanisms. As our vision is to use BP as a unifying modeling language suited for combining these methods, we demonstrate the interplay of the different mechanisms described in the preceding sections. Specifically, we show how the integrations with SMT solvers, probabilistic modeling, and DRL can be used together to create a comprehensive environment that facilitates both modeling and analysis.

For this demonstration, we use a variation of the \emph{bit-flip} problem described earlier: A two-player game over a Boolean matrix of dimensions $N\times M$. The initial state of the matrix resembles a chessboard with alternating positive and negative bits. Our opponent randomly flips rows and columns, while our objective is to devise a strategy that involves strategically flipping rows or columns to turn on as many bits as possible \emph{simultaneously}. We can randomly flip bits or implement a greedy strategy that selects the row or column with the highest number of bits to turn on in each round. However, as we will demonstrate later, a better strategy that is more challenging to implement manually can be achieved using DRL.

\autoref{sec:smt} and \autoref{sec:probabilistic-modeling} demonstrate how the discrete implementation of the problem can be challenging for execution and analysis as the matrix size increases. This section uses a solver-based implementation combined with the \li{choice} idiom presented in \autoref{sec:probabilistic-modeling} to model the opponent's behavior. The complete program used for the following evaluation, along with a description of the code, is available in an appendix in the supplementary material.

We ran the PPO algorithm~\cite{huang_closer_2022} to learn a strategy for the task. The learned strategy was compared against two baseline strategies. The first strategy randomly selects rows and columns to flip, while the second greedily selects the row/column with the highest number of bits to turn on in each round. The mean expected reward of these strategies was computed using the \li{choice} idiom. Results are presented in \autoref{fig:bit-flip-drl}. We observe that the PPO algorithm rapidly obtained a strategy that can plan ahead and achieve better (or equal in the $3 \times 3$ case) results than the greedy approach in all matrix sizes examined.

The bit-flip demonstration shows how BP can serve as a ``Swiss army knife'' for modeling and analysis: Starting with a problem that was challenging to execute in a discrete program and upgrading it to utilize solvers. Subsequently, DRL can be applied to achieve a more effective strategy, and it can be evaluated and compared using probabilistic analysis.

The integration of SMT, probabilistic model checking, DRL, and BP holds great potential, as the first three are already closely linked. Many probabilistic model checking problems are solved using SMT-based constraint solving or linear optimization, while DRL algorithms address scalability issues in these problems. Tools like PRISM and STORM demonstrate this synergy, combining solvers and DRL. Extending this integration to BP could enable adaptive decision-making with formal guarantees for complex systems.

\vspace{-10pt}

\section{Conclusion}
\vspace{-3pt}

In conclusion, our study focused on integrating BP with a variety of techniques to establish a comprehensive framework for specifying and analyzing reactive systems. Moving forward, our future work will delve deeper into use cases involving multiple integrations and explore more complex and extensive examples. Additionally, we plan to focus on evaluating the usability of our proposed framework and assess how well it aids in programming to increase adoption and enhance accessibility.

\vspace{-10pt}
\section*{\uppercase{Acknowledgements}}
\vspace{-3pt}
This work of Weiss, Yaacov, and Zisser was partially supported by funding from the Israel Science Foundation (ISF) grant number 2714/19. The work of Ashrov and Katz was partially funded by the European Union (ERC, VeriDeL, 101112713). Views and opinions expressed are however those of the author(s) only and do not necessarily reflect those of the European Union or the European Research Council Executive Agency. Neither the European Union nor the granting authority can be held responsible for them.






\bibliographystyle{apalike}
{\small
\bibliography{reduced_short.bib}}

\end{document}